\documentclass[preprint,11pt]{article} 

\usepackage{ol2}
\usepackage{epstopdf}
\usepackage{color, graphicx, subfigure}
\usepackage{mathptmx}
\usepackage{bm}
\usepackage{times}
\usepackage{cite}
\usepackage{xcolor}
\newcommand\Eint{\mathbf{E}^\mathrm{int}}
\newcommand\Esca{\mathbf{E}^\mathrm{sca}}
\newcommand\Ei{\mathbf{E}^\mathrm{i}}
\newcommand\Ajmy{\mathbf{A}_{jm_z}^{(y)}}

\newcommand\nal{n_{\mathrm{Al}_2\mathrm{O}_3}}
\newcommand\nsi{n_{\mathrm{SiO}_2}}

\begin{document}

\title{Dual and anti-dual modes in dielectric spheres}

\author{Xavier Zambrana-Puyalto$^{1,2}$, Xavier Vidal$^{1}$, Mathieu L. Juan$^{1,2}$ and Gabriel Molina-Terriza$^{1,2}$}

\address{$^{1}$Department of Physics and Astronomy, Macquarie University, 2109 NSW, Australia \\ $^{2}$ARC Centre for Engineered Quantum Systems, Macquarie University, 2109 NSW, Australia}

\email{xavier.zambranapuyalto@mq.edu.au} 

\begin{abstract}
We present how the angular momentum of light can play an important role to induce a dual or anti-dual behaviour on a dielectric particle. Although the material the particle is made of is not dual, i.e. a dielectric does not interact with an electrical field in the same way as it does with a magnetic one, a spherical particle can behave as a dual system when the correct excitation beam is chosen. We study the conditions under which this induced dual or anti-dual behaviour can be induced.
\end{abstract}


\pagebreak

\section{Introduction}

One of the most intriguing mysteries of modern physics is the principle of charge quantization. As Dirac proved in 1931, the existence of only one magnetic monopole would be sufficient for all the electric charges to be multiple of a certain value \cite{Dirac1931}. Nevertheless, magnetic monopoles are yet to be found \cite{Adrian-Martinez2012,Abbasi2013}. This fact has a very important implication in electromagnetism and quantum electrodynamics: Maxwell equations are not symmetric with respect to electric and magnetic fields. In contrast, it is well known that Maxwell equations in free space are symmetric under duality transformations \cite{Jackson1998}. This transformation mixes the electric and magnetic fields through a continuously varying parameter. Therefore, we can define a generator of this transformation. Indeed, in 1965 Calkin found that the helicity of a light beam is the generator of duality transformations \cite{Calkin1965}. However, the fact that duality symmetry is always broken for material media has mitigated the use of the helicity of light to probe light-matter interactions. In this research line, a new finding was presented very recently in \cite{preIvan2012}. It was proven that the macroscopic Maxwell equations for isotropic and homogeneous media can be dual-symmetric if some conditions are fulfilled. Microscopically, duality symmetry is still broken, but the collective effect of all the charges and currents in the medium restores the symmetry in the macroscopic approximation. In this work, different samples were probed and the non-conservation of helicity was carefully quantified. In the same way as it happens with any other generator of symmetries, if the helicity of a light beam is preserved upon interaction with a material medium, this necessarily implies that the system is symmetric under its associated duality symmetry. We refer to these sort of media as 'dual'.

In this paper, we propose a method to effectively convert a non-dual arbitrarily large dielectric sphere into a dual particle. This means that if we probe the system with light beams whose value of the helicity is well defined, the helicity of these light beams will be preserved upon interaction. Our method to restore duality is based on an analytical description of spheres in terms of Mie coefficients and multipolar modes. Using a method to control the scattered field introduced in \cite{Zambrana2012}, we are able to effectively induce duality symmetry on the particle, regardless of its size and index of refraction. Recently, dielectric particles are starting to gather a lot of interest in metamaterial sciences \cite{Zhao2009}. Their lack of losses, their directional properties \cite{Nieto-Vesperinas2011,Liu2012ACS} and their ability of induce both electric and magnetic dipoles \cite{Garcia-Etxarri2011, Evlyukhin2010} are thought to be applicable not only in metamaterials, but also in nanophotonics and stealth technology. Indeed, it has been proven in recent experiments that the so-called first Kerker condition of zero backward scattering can be achieved both in the microwave and optical regime \cite{Geffrin2012,Person2013,Fu2013}. Also, it has been shown in \cite{Zambrana2013} that a dual and cylindrically symmetric system has zero backscattering, whereas an anti-dual and cylindrically symmetric system has zero forward scattering, although this last one could only be achieved with active particles \cite{Garcia-Camara2011,Nieto-Vesperinas2011,Alu2010} 

\section{Helicity and Generalized Lorenz-Mie Theory}
In this section, we will provide the reader with the basic concepts and formulae necessary to understand the methods used in the forthcoming sections. To begin with, we will introduce the generator of duality transformations - the helicity. The helicity is defined as the projection of the angular momentum (AM) onto the normalized linear momentum, \textit{i.e.} $\Lambda={\mathbf{J}}\cdot{\mathbf{P}}/|{\mathbf{P}}|$ \cite{Tung1985,Lifshitz1982,Ivan2012PRA}. It can also be expressed for monochromatic fields as a differential operator: $\Lambda = \vert k \vert^{-1}(\nabla \times )$ \cite{Messiah1999}. It has two eigenvectors with respective eigenvalues $p = \pm 1$ and it commutes with all the components of the linear and AM operators, $\mathbf{P}$ and $\mathbf{J}$.  
Finally, in the Fourier space, the helicity measures the handedness in all the plane waves. If all the plane waves have the same circular polarization with respect to their own propagation direction, then the beam will have a well defined helicity, otherwise it will not. 

Now, since we will be working with the scattering of spheres, we will use the Generalized Lorenz-Mie Theory (GLMT) to solve the scattering problem \cite{GLMT_book}. The GLMT solves the interaction between an arbitrary incident EM field propagating in a lossless, homogeneous, isotropic medium and a homogeneous isotropic sphere. The problem is described with three EM fields: the incident ($\Ei$) on one hand, and the scattered ($\Esca$) as well as the interior ($\Eint$) on the other hand. The three fields in the problem are decomposed into multipolar modes $\Ajmy$ and then the boundary conditions are applied. The multipolar modes are a complete basis of Maxwell equations and they are particularly suitable for problems with spherical symmetry. They are eigenvectors of the total AM operator $J^2$ and one of its projections such as $J_z$, with respective values $j$ and $m_z$ \cite{Rose1955,Jackson1998}. Furthermore, they are eigenvectors of the parity operator $\Pi$ with values $(y)=(m)$ for a $(-1)^j$ parity, and $(y)=(e)$ for $(-1)^{j+1}$, where $(m)$ and $(e)$ stand for magnetic and electric multipole. 

We will excite the dielectric spheres with cylindrically symmetric beams, \textit{i.e} excitation beams $\Ei$ with a well defined value of the z component of the AM, $J_z$. Moreover, we will also want our beams to have a well-defined helicity value, so that it can be easily characterized if the particle is dual or not by computing the helicity transfer from the incident component to the opposite one. The decomposition of these beams into multipoles can be done analytically when they are paraxial \cite{Zambrana2012JQSRT}, or semi-analytically in the general case \cite{Gabi2008,Zambrana2012}. Once the decomposition of the incident beam is found, the expression of $\Esca$ and $\Eint$ is given by the GLMT. Actually, the formal expression of $\Esca$ and $\Eint$ is almost the same one as $\Ei$. The only difference comes from the fact that each of the multipolar modes is modulated by a coefficient that depends on its AM and parity. These are the so-called Mie coefficients \cite{Bohren1983}. Also, it is worth noticing that in order to fulfil the boundary conditions the radial functions in the multipolar modes of the scattered field must be Hankel functions, while in both the incident and interior fields are Bessel functions \cite{Bohren1983}. The general expression for the three fields in the problem is the following one:
\begin{equation}
\begin{array}{ccl}
\Ei &= &\displaystyle\sum_{j=\vert m_z \vert}^{\infty} i^j (2j+1)^{1/2}  C_{jm_zp} \left[ \textbf{A}_{jm_z}^{(m)}+ip\textbf{A}_{jm_z}^{(e)} \right]  \\
\Esca &=&\displaystyle\sum_{j=\vert m_z \vert}^{\infty} i^j (2j+1)^{1/2}  C_{jm_zp} \left[ b_j\textbf{A}_{jm_z}^{(m)}+ipa_j\textbf{A}_{jm_z}^{(e)} \right]  \\
\Eint &=&\displaystyle\sum_{j=\vert m_z \vert}^{\infty} i^j (2j+1)^{1/2} C_{jm_zp} \left[ c_j\textbf{A}_{jm_z}^{(m)}+ipd_j\textbf{A}_{jm_z}^{(e)} \right]
\end{array}
\label{fields}
\end{equation}
\normalsize
where $p= \pm 1$ is the helicity of the incident beam, $m_z$ is the angular momentum projection on the $z$ axis of the incident beam, and $C_{jm_zp}$ is a function that modulates the multipolar content of the incident field. As such, $C_{jm_zp}$ is a function of the incident beam properties: $m_z$ and $p$, but also its transversal momentum profile. The mathematical expression for $C_{jm_zp}$ is given in \cite{Zambrana2012}. The Mie coefficients $\left\lbrace a_j,b_j,c_j,d_j\right\rbrace$ only depend on the size parameter of the problem ($x=2 \pi r/\lambda$), the relative permeability ($\mu_r$) and permittivity ($\epsilon_r$) of the sphere with respect to the surrounding medium. Here, $r$ is the radius of the particle and $\lambda$ the wavelength in free space. 

Now, as it has been discussed and proven in \cite{Zambrana2013}, it is crucial to note that even though the incident field has a well-defined helicity, the scattered and interior fields do not generally have it. This is a consequence of the fact that the two pairs of Mie coefficients $\left\lbrace a_j,b_j \right\rbrace$ and $\left\lbrace c_j,d_j \right\rbrace$ are not generally equal. In fact, we can split the total energy of the scattered field ($w^\mathrm{sca}$) on a control sphere into two parts. These two parts account for the energy scattered in modes with the same helicity as the incident field ($w_p^\mathrm{sca}$) and with the opposite helicity ($w_{-p}^\mathrm{sca}$):
\begin{eqnarray}
w^\mathrm{sca} & = & \sum_j  (2j+1)\vert C_{jm_zp} \vert^2 \left( \vert a_j \vert^2 + \vert b_j \vert^2\right) \\ 
\label{dual}
w_p^\mathrm{sca} & = & \sum_j (2j+1)\vert C_{jm_zp} \vert^2 \vert a_j +   b_j \vert^2 \\
w_{-p}^\mathrm{sca} & = & \sum_j (2j+1)\vert C_{jm_zp} \vert^2 \vert a_j -   b_j \vert^2  
\label{antidual}
\end{eqnarray} 
Note that in general the scattered field will always carry energy in modes with the opposite helicity, \textit{i.e.} $w_{p-}^{sca} \neq 0$. However, if $a_j(x)=b_j(x) \  \forall j$ the particle only scatters energy with the same helicity as the incident beam, conserving the helicity of the electromagnetic (EM) field and therefore behaving as a dual medium. Note also that if $a_j(x)=-b_j(x) \ \ \forall j$, then $w_p^{sca}=0$. That is, the scattered field has the opposite helicity to the incident one. We refer to such scatterers as anti-dual \cite{Zambrana2013}. It can be proven that a spherical particle will only be dual if $\mu_r=\epsilon_r$. Also, as it has been stated in the introduction, it has been proven that a dielectric material cannot be anti-dual \cite{Nieto-Vesperinas2011,Alu2010} and that such materials could only be made of active media \cite{Garcia-Camara2011}. These two facts have been experimentally verified in the dipolar approximation \cite{Geffrin2012,Person2013,Fu2013}, experimentally achieving $a_1=b_1$ and $a_1 \approx -b_1$. Below we show how to extend those conditions to other regimes and also we show that there are situations where one can even approximately fulfil the anti-dual condition.

\section{Induced duality symmetry}
In this section we propose a method to induce duality symmetry on dielectric spheres. Our method is based on the following idea. Suppose that we have a single isotropic and homogeneous dielectric sphere with a given size $r$ and index of refraction relative to the surrounding medium $n_r$. The magnetic permeability can be set to 1 for simplicity. We will describe the EM response of the sphere with the GLMT. Now let's suppose that the behaviour of the particle could be described by only two Mie coefficients of the same order $a_n$ and $b_n$ for a certain range of the size parameter $x$. If in that range $a_n$ is equal to $b_n$, then the particle will be dual \cite{Kerker1983,preIvan2012, Zambrana2013}. Nevertheless, the EM response of a particle cannot usually be only described with two Mie coefficients. And even when that is the case, it is not clear that $a_n$ can be equal to $b_n$ in that regime. For example, in the dipolar or Rayleigh approximation \cite{vandeHulst1957, Kerker1983}, the EM response of the particle can be described with $a_1$ and $b_1$. Nonetheless, the dipolar approximation is only valid for $x\ll 1$. That is, if we fix the size of the particle, the approximation will only be valid for wavelengths such that $\lambda \gg 2\pi r$. In this regime, the electric dipolar moment is typically much larger than the magnetic one. Then, if there are certain values of the parameters where $a_1(x_1, n_r)=b_1(x_1, n_r)$, we would still have to validate that all the higher multipolar moments are small. 

\begin{figure}[tbp]
\centering\includegraphics[width=9cm]{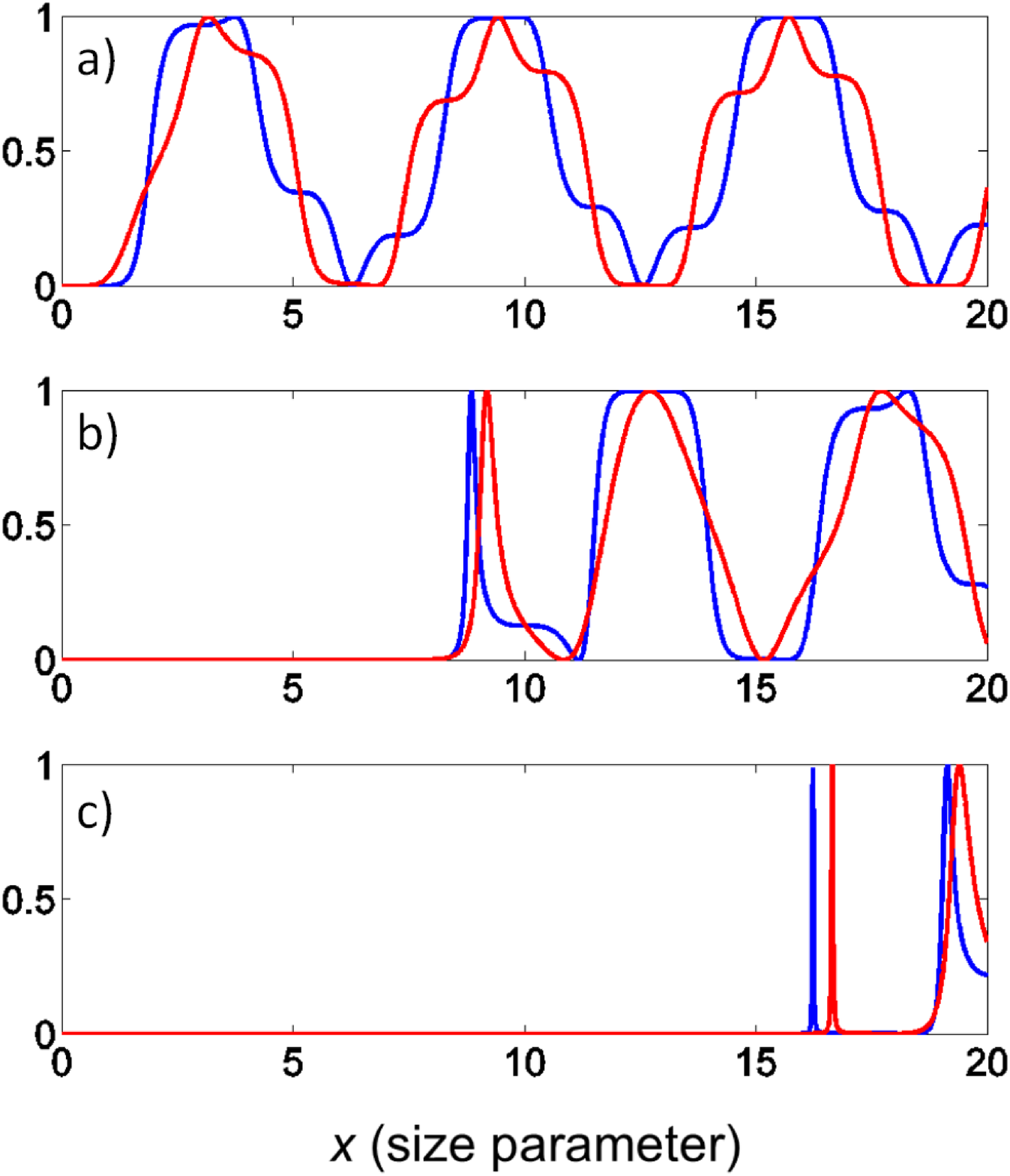}
\centering\caption{Norm of the Mie coefficients $\vert a_n \vert $ (in red) and $\vert b_n \vert$ (in blue) for $n=1,10,20$ in a), b) and c), respectively as a function of the size parameter $x=2\pi r/\lambda$ . The relative index of refraction is $n_r=1.5$. It can be seen that all of them start being significantly different from zero when $x \approx 4n/5$. \label{ajbj}}
\end{figure}

In this section, we will show that by using cylindrically symmetric modes, a particle can be described with only two Mie coefficients ($a_n$ and $b_n$) for certain regimes. Also, we will show that the duality condition $a_n=b_n$ can be achieved for arbitrary large $n$. This duality condition is achieved in three steps. In the first place, the lower $n-1$ Mie coefficients are not excited. Then, the excitation wavelength is chosen so that $a_n=b_n$. Finally, we only choose the situations where the helicity change due to the higher order Mie coefficients is negligible. 

In order to fulfill this program, we will first describe the behaviour of the Mie coefficients for dielectric particles, \textit{i.e.} when the relative index of refraction $n_r$ is real, in order to get a deeper understanding of the phenomena involved. In this case, the Mie coefficients, $a_j$ and $b_j$, that is the multipolar moments of the sphere, are complex and their absolute values are bounded between zero and one, $0\leq \left\lbrace \vert a_j\vert,\vert b_j\vert \right\rbrace \leq 1$. A multipolar moment of order $n$ is very close to zero for small $x$, and they start to grow for a value of $x$ which is proportional to the order of the mode $n$. The proportionality value depends on $n_r$. For example, when $n_r=1.5$, it can be computationally verified that the Mie coefficients approximately start to grow when $x \approx 4n/5$ (see Fig. 1). Then, if $x < 4n/5$ the multipolar moments of order $n$ are negligible. For $x > 4n/5$ their absolute values oscillate between 0 and 1. Hence, on one hand, it is always true that it exists an interval around $x \approx 4n/5 $ where $a_n$ and $b_n$ start growing and the higher Mie coefficients are approximately zero (as they start growing for $x \approx 4(n+1)/5$). But on the other hand, it is impossible that all the first $n-1$ Mie coefficients are zero when $a_n$ and $b_n$ are not, as they start growing for smaller $x$. This is depicted in Fig. 1, where $a_n$ and $b_n$ are plotted for $n=1,10,20$. It can be observed that all the Mie coefficients follow the pattern described above: their absolute value is 0 until $x \approx 4n/5$, and then they oscillate between 0 and 1. 

Because of this behaviour, it would seem that condition $a_n=b_n$ cannot be met. However, it has been shown in \cite{Zambrana2012} that the first $n-1$ Mie coefficients (and its associated multipolar modes) can be removed from the scattered field when a beam with $\vert J_z \vert =n$ is used to illuminate a sphere. This is a consequence of the conservation rules for the AM. Thus, we can isolate an arbitrary pair of Mie coefficients $a_n$ and $b_n$ around $x \approx 4n/5$: the first $n-1$ Mie modes can be removed from the scattering using a beam with $\vert J_z \vert =n$, and the higher modes are naturally attenuated. The only remaining point that needs to be discussed is the helicity change induced by the higher order modes. As we have stated previously, if a light beam with $\vert J_z \vert =n$ is used, $a_n$ and $b_n$ are going to be the dominant Mie modes around $x \approx 4n/5$. However, generally the condition $a_n(x_n^*,n_r)=b_n(x_n^*,n_r)$ will be met for a particle such that $x_n^* > 4n/5$. If this value $x_n^*$ is close to $4(n+1)/5$, the Mie coefficients $a_{n+1}$ and $b_{n+1}$ cannot be ignored. Thus, as stated above, what needs to be carefully studied is the helicity change induced by $a_{n+1}$, $b_{n+1}$ and the higher orders when the condition $a_n=b_n$ is met. This is studied in the next subsections for spheres of different sizes and materials. The wavelength will be fixed at $\lambda=780$nm unless the contrary is stated.

\subsection{Gaussian excitation}
\begin{figure}[tbp]
\centering\includegraphics[width=14cm]{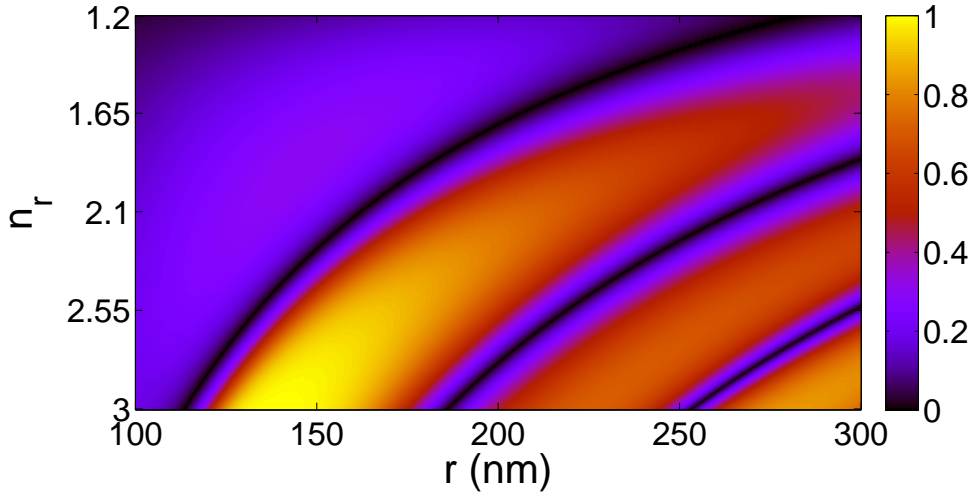}
\caption{Plot of $ \vert a_1-b_1 \vert $ as a function of the radius of the sphere $r$ (horizontal axis) and the relative index of refraction $n_r$ (vertical axis). It can be observed that there are three major regions where $ \vert a_1-b_1 \vert =0$. \label{a1b1}}
\end{figure}
In this subsection, our method to induce duality symmetry in a dielectric sphere is tested with a Gaussian beam. Nonetheless, in order to get some intuition, we first plot  $\vert a_1-b_1 \vert$ as a function of the radius of the particle $r$, and the relative index of refraction $n_r$. This is depicted in Fig. 2. It can be observed that for any value of the refractive index, multiple radius of the sphere satisfy $a_1 = b_1 $. This is a consequence of the oscillating behaviour of the Mie coefficients seen in Fig. 1. It is also interesting to note that the larger the refractive index is, the smaller the particle is when $a_1 = b_1$. Nonetheless, looking at Fig. 2, one does not know how good the approximation of only describing the sphere with the $a_1$ and $b_1$ is - we are missing all the information due to the higher modes. To capture this behaviour, we define the transfer function $T_{{m_z}p}(r,n_r)$: 
\begin{equation}
T_{m_zp}(r,n_r)=  \frac{w_{-p}^\mathrm{sca} }{w_{p}^\mathrm{sca} } = \frac{\sum_{j=m_z  }^{\infty} (2j+1)\vert C_{jm_zp} \vert^2  \vert a_{j}  -  b_{j} \vert^2  } {\sum_{j=m_z  }^{\infty} (2j+1)\vert C_{jm_zp} \vert^2 \vert a_{j}  +  b_{j} \vert^2 } 
\label{Tj}
\end{equation}
This function gives the fraction of scattered light going to modes with opposite helicity with respect to the incident light ($w_{-p}^\mathrm{sca}$), compared with the fraction of scattered light going to modes with the same helicity ($w_{p}^\mathrm{sca}$), for a given angular momentum ($m_z$) and helicity ($p$) of the incident beam. Hence, $T_{m_zp}(r,n_r)$ varies from 0 to infinity. When $T_{{m_z}p}(r,n_r)$ tends to zero, the particle is dual and all the scattered light has helicity $p$: in other words, it fulfils the first generalized Kerker condition \cite{Zambrana2013}. On the contrary, when $T_{m_zp}(r,n_r)$ tends to infinity, the particle is anti-dual and all the scattering is transferred to the cross helicity, $-p$: it fulfils the generalized second Kerker condition \cite{Zambrana2013}.

\begin{figure}[tbp]
\centering\includegraphics[width=14cm]{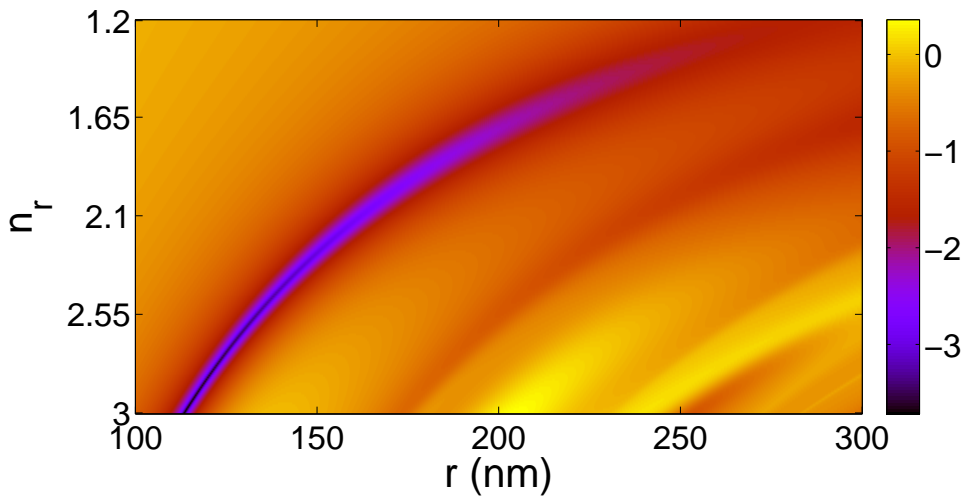}
\caption{ Plot of the $log$ of the Transfer function $T_{m_zp}(r,n_r)$ for $m_z=1$ and $p=1$, \textit{i.e.} $\log\left( T_{11}(r,n_r)\right) $, as a function of the radius $r$ of the particle (horizontal axis) and the relative index of refraction $n_r$ (vertical axis). The radius of the particle is varied from 100nm (left) to 300nm (right) and the relative index of refraction goes from $1.2 \leq n_r \leq 3$. \label{T11}}
\end{figure}

Figure \ref{T11} shows the value of $\log\left( T_{11}(r,n_r)\right) $, i.e. for a focused Gaussian beam with well defined helicity. We use the same range of parameters, $\left\lbrace r,n_r \right\rbrace$, used in Fig. \ref{a1b1}. The logarithm is applied to stand out the dual behaviour of the particle. It can be observed that now there is only one region in the $(r,n_r)$ space where the dual condition is fulfilled, in contrast to Fig. \ref{a1b1} where three different regions had $\vert a_1 - b_1 \vert = 0$. This fact was expected due to the behaviour of the Mie coefficients explained in Fig. \ref{ajbj}. Indeed, $a_2$, $b_2$ and the higher orders are no longer negligible for large values of $r$ and therefore they induce a helicity transfer from $p$ to $-p$. Furthermore, it is interesting to see how the induced duality strongly depends on the relative index of refraction. It is apparent from Fig. \ref{T11} that the dielectric sphere gets closer to the dual condition when $n_r$ gets larger. That means that the helicity of the incoming beam will be better preserved when particles with a high refractive index embedded in a low refractive index medium are used.
In addition to the duality considerations mentioned above, Fig. 3 also depicts the anti-dual behaviour of the particle. Two main features can be observed. First, as it can be deduced from the colorbar, the anti-dual condition is not achieved as finely as the dual is. That is, the generalized second Kerker condition is more difficult to achieve. Hence, the forward scattering is never reduced as much as the backward is. This is consistent with the few experiments done until now \cite{Geffrin2012,Fu2013} and with the fact that dielectric particles cannot be anti-dual \cite{Nieto-Vesperinas2011,Alu2010,Garcia-Camara2011}. Secondly, if the relative index of refraction is maintained, the anti-dual condition is held for larger particles than the dual one.   

\begin{figure}[tbp]
\centering\includegraphics[width=2.5in]{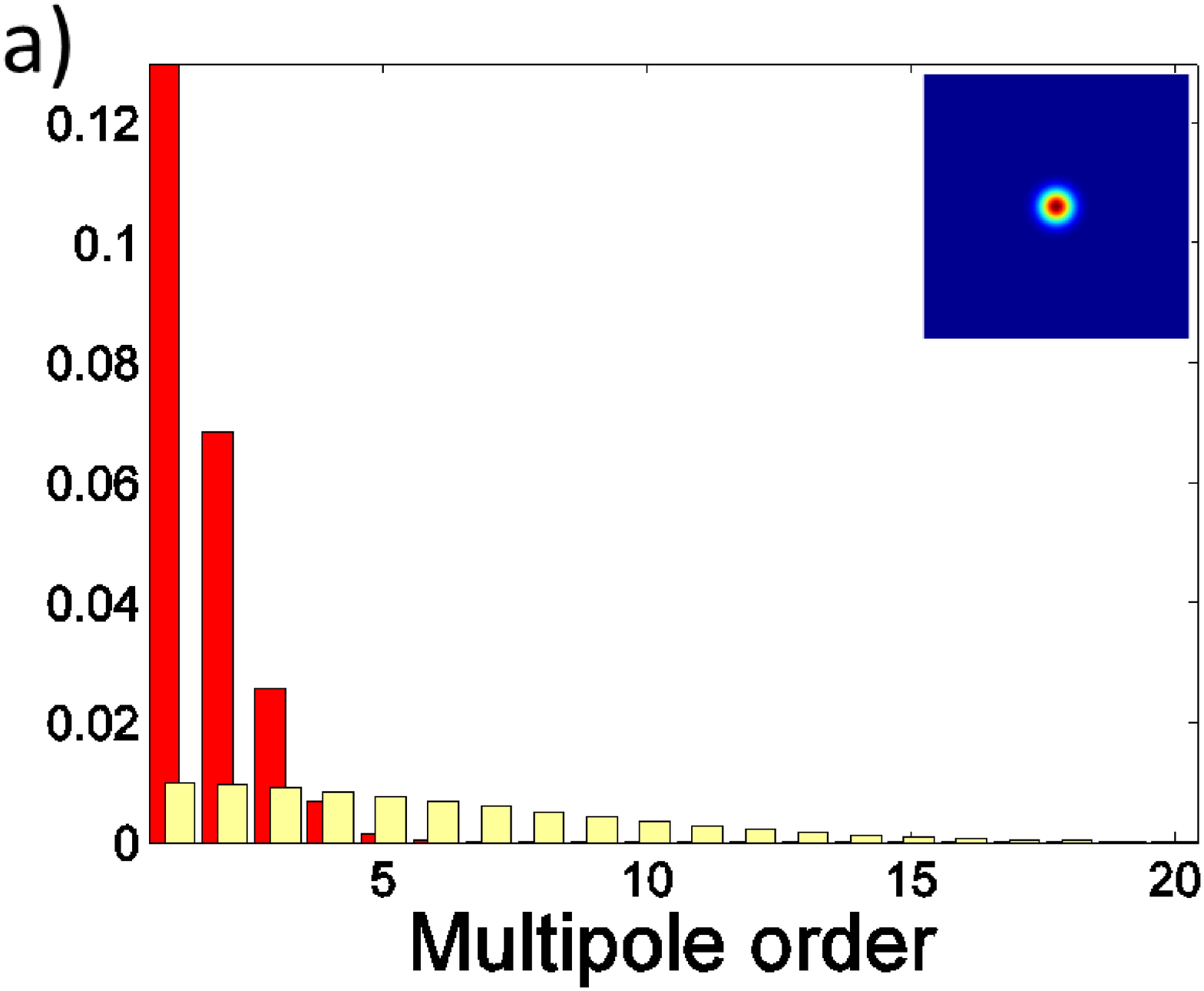} \includegraphics[width=2.5in]{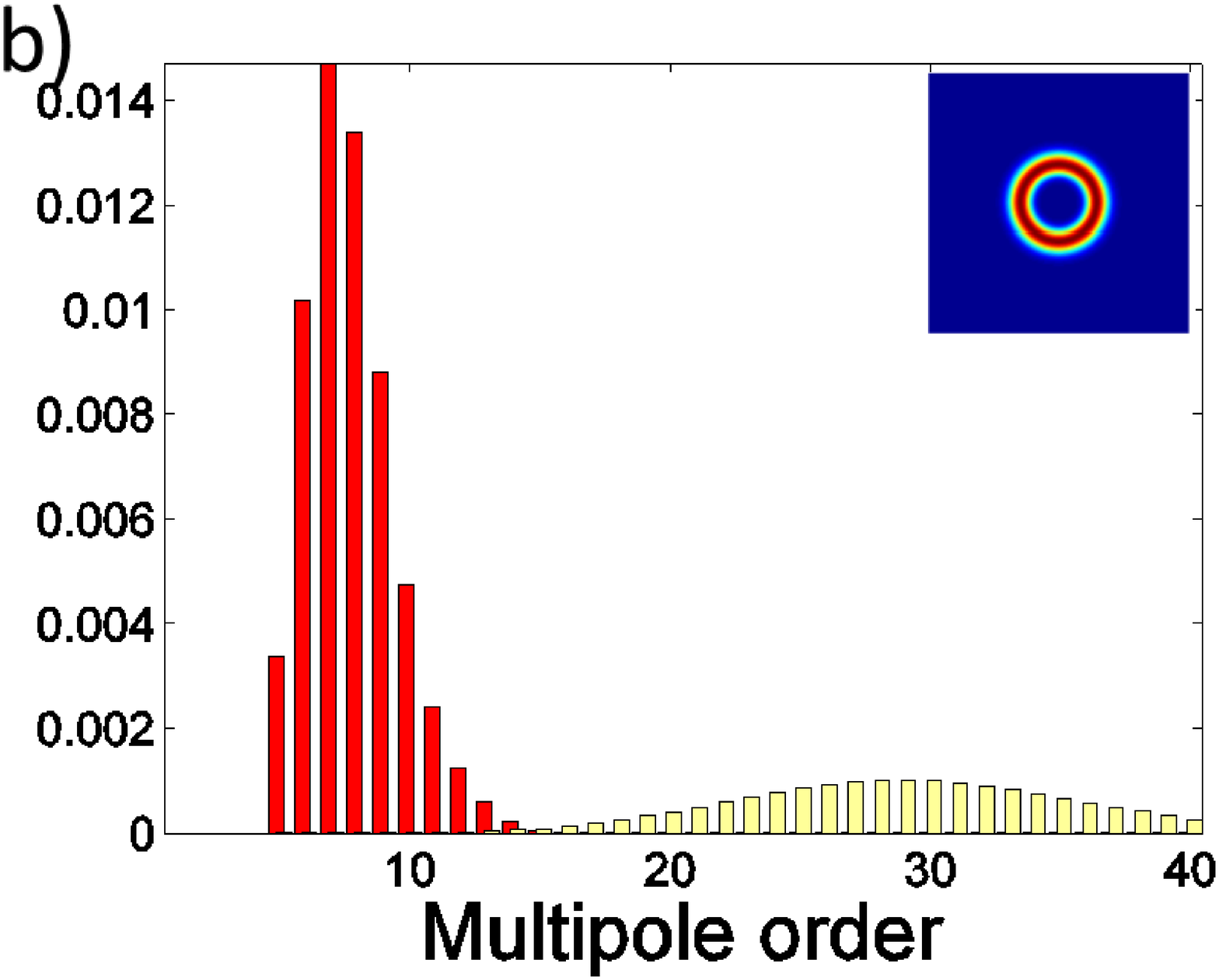}
\caption{Multipolar decomposition ($\vert C_{jm_zp} $) for the two different incident beams used in this article. The function $C_{jm_zp}$ is normalized with the following relation: $\sum_j(2j+1)\vert C_{jm_zp} \vert^2=1$ \cite{Gabi2008}. The insets represent the intensity plots of the modes used for each simulation. The yellow coloured bars indicate NA=0.25, and the red ones NA=0.9. The multipolar decomposition of (a) Gaussian beam and (b) LG$_{0,4}$ is presented. In both cases, the helicity is chosen to be $p=1$. $\vert C_{jm_zp} \vert$ is plotted in the \textit{y} axis and the multipolar order $j$ is plotted in the \textit{x} axis.  \label{Cjmzp}}
\end{figure}

Last but not least, there is a subtlety in Eq. (\ref{Tj}) that needs to be commented. The transfer function $T_{m_zp}(r,n_r)$ depends on the incident beam $\Ei$, through the multipolar expansion $C_{jm_zp}$. As mentioned earlier, this will depend not only on the eigenvalues $m_z$ and $p$, but also on its transversal properties. In particular, the value of the amplitudes $C_{jm_zp}$ will vary depending on how much the beam is focused \cite{Zambrana2012,Mojarad2008,Mojarad2009}. 
An example is provided by Fig. \ref{Cjmzp}. In Fig. \ref{Cjmzp}(a), a circularly polarized Gaussian beam is focused with two different numerical apertures (NA). In yellow, the beam is focused with a NA$=0.25$, and in red a lens of NA$=0.9$ is used. In Fig. 4(b), a Laguerre-Gaussian beam with radial number $q=0$ and azimuthal number $l=4$ (LG$_{0,4}$) \cite{prePampaloni2004} is depicted, and same colours are used regarding the focusing strength. It can be seen that the multipolar decomposition is narrowed down when the NA of the lens is increased \cite{Zambrana2012}. Nevertheless, after carrying out many simulations, we have realized that, even though  $T_{m_zp}(r,n_r)$ will be different for each particular case, the qualitative behaviour that we describe will not change appreciably with the NA of the focusing lens. Hence, we will use the same NA$=0.9$ for the rest of the article.

\subsection{Higher AM modes excitation}
Now, we will show how the dual and anti-dual properties of the particle will dramatically change when using a higher angular momentum mode as an incident field. A cylindrically symmetric beam with a well defined helicity and $m_z=5$ is used. In particular, the beam is a LG$_{0,4}$ focused with a lens whose NA$=0.9$. Its decomposition into multipoles is given in Fig. 4(b). As we have discussed before, if we excite a sphere with an eigenvector of $J_z$ with eigenvalue $m_z$, we can describe the EM response of it with $a_{\vert m_z \vert}$ and $b_{\vert m_z \vert}$ as the ${\vert m_z \vert}-1$ first Mie coefficients do not contribute to the scattering. However, as we did in the previous subsection, the helicity change induced by the Mie modes whose order is higher than $n=5$ has to be carefully studied. 

\begin{figure}[tbp]
\centering
\includegraphics[width=2.5in]{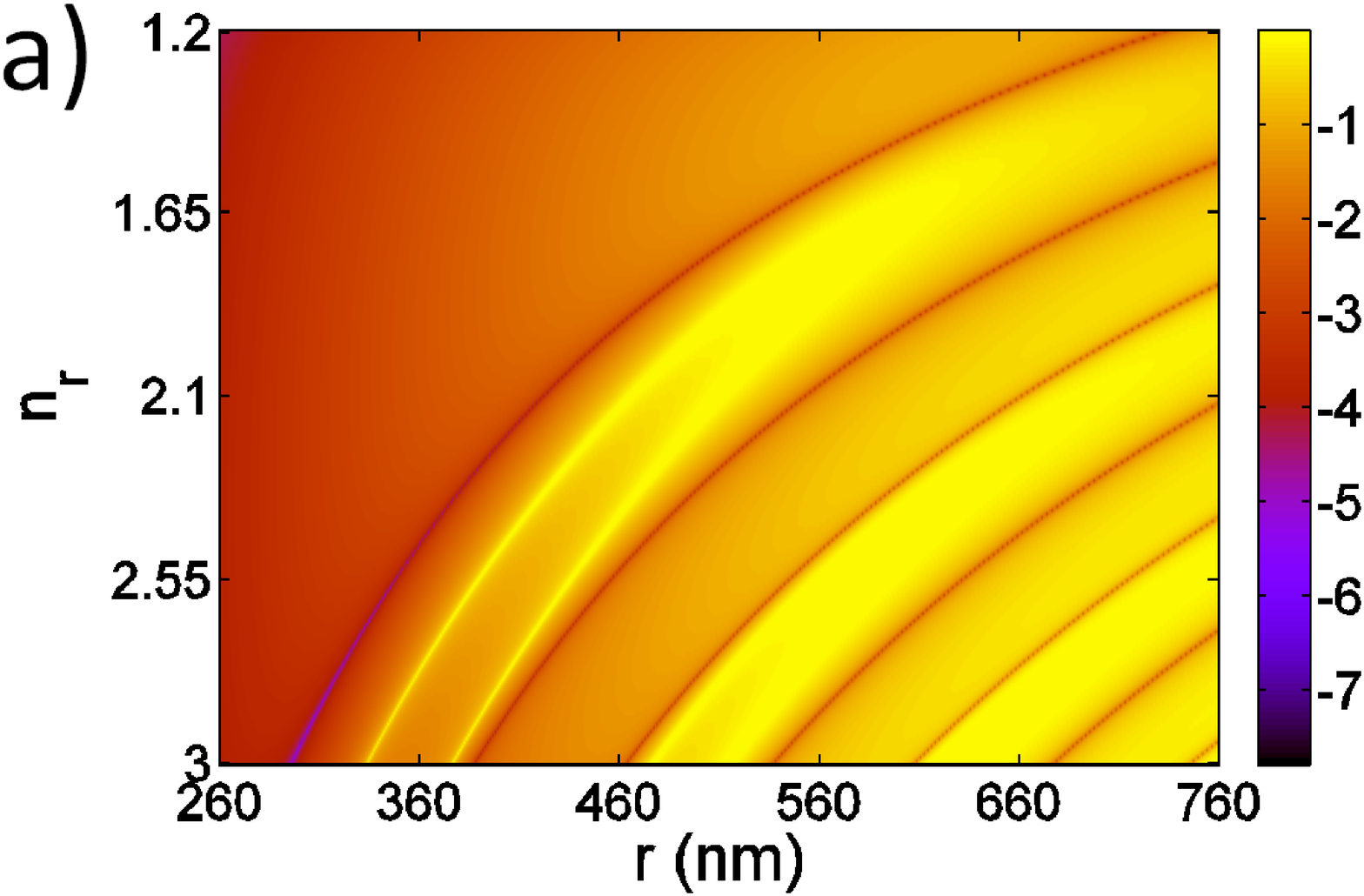} \includegraphics[width=2.5in]{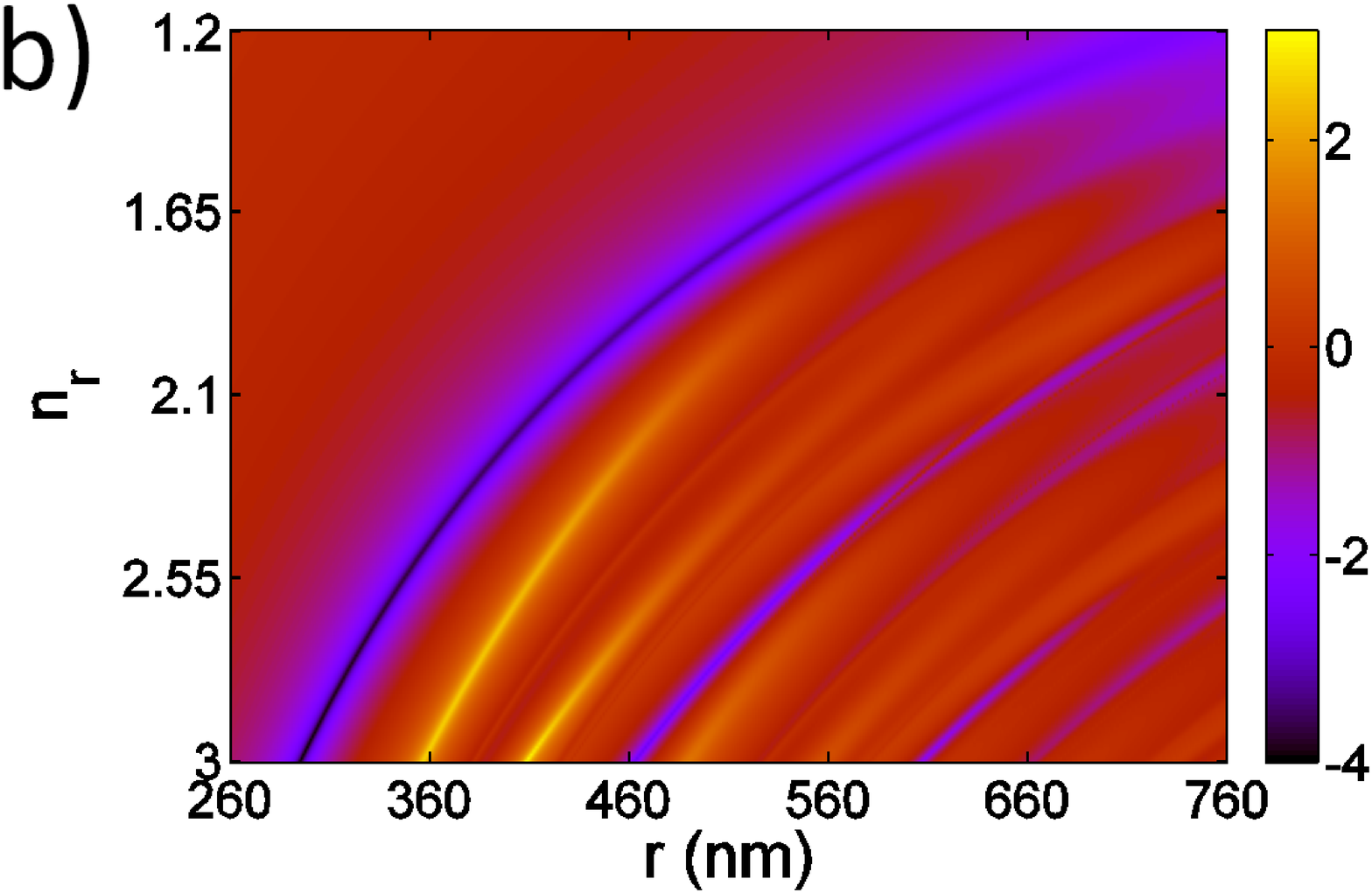} 
\caption{ a) Plot of $\log \left( \vert a_5-b_5 \vert \right) )$ as a function of the radius $r$ of the particle (horizontal axis) and the relative index of refraction $n_r$ (vertical axis). b) Plot of the $log$ of the transfer function $T_{m_zp}(r,n_r)$ for $m_z=5$ and $p=1$, $\log \left( T_{51}(r,n_r) \right) $, as a function of $r$ and $n_r$. \label{a5b5}}
\end{figure}

In Fig. \ref{a5b5}(a), it can be observed the shape of $\log \left( \vert a_5-b_5 \vert \right)$, where the $\log$ function has been used to make the plot more intelligible. Although Fig. 5(a) has a similar shape to $ \vert a_1-b_1 \vert $, some differences can be found. The range of sizes for the particles to achieve the dual condition $a_5=b_5$ has increased for the same interval of $n_r=[1.2,3]$. Before, it spanned 200nm, whereas now it spans 500nm. As previously discussed, though, this plot does not enables us to see how dual-symmetric (or anti-dual) the particle is. Therefore, we have plotted $\log \left( T_{51}(r,n_r) \right)$ in Fig. 5(b) to capture this behaviour. As in the $j=1$ case, an increase in $n_r$ is linked to an increase of the dual properties of the sphere. Also, it can be inferred from a comparison between  Fig. 5(b) and Fig. 3, the dual and anti-dual conditions are fulfilled with a better approximation in this new occasion. Indeed, the minimum value of the colorbar drops almost an order of magnitude, and the maximum value increases more than two orders of magnitude. That is, for some certain combinations of $r$ and $n_r$, the energy of the scattered field in the modes of opposite helicity is $10,000$ times smaller than the energy going to the original helicity of the incident field; whereas for some other certain conditions, the scattered field energy is dominated by modes with the opposite helicity with a ratio of 500 to 1. As in the incident Gaussian beam case, an increase in $n_r$ is linked to an increase of the dual and anti-dual properties of the sphere. To summarize, some very general conclusions can be reached after a careful look into these results:
\begin{itemize}
\item The larger the relative refraction index of the particle $n_r$ is, the more accurately the two generalized Kerker conditions can be achieved.
\item For a constant $n_r$, the bigger the particle is, the more dual (and anti-dual) the particle can become if high AM modes at the adequate wavelength are used to excite it.
\item Fixing the index of refraction and size of the particle, we can always approximately induce the duality symmetry by choosing a right combination of AM and optical frequency of our laser, regardless of how big the sphere is with respect to the wavelength.
\end{itemize}

We have confirmed these conclusions by doing calculations similar to the ones presented with different sizes, wavelengths, refractive indexes, and excitation beams. The results are always consistent with the conclusion above. Finally, note that in our simulations $n_r \leq 3$ and $\lambda = 780$nm. With these two conditions, the smallest a particle can be to induce duality symmetry is 120nm. This size could be reduced down to 81nm if a $\lambda=532$nm was used. 

\section{Proposals for experimental implementations}
In the previous section, we have compared different scenarios where the EM duality symmetry can be induced with dielectric spheres and arbitrarily high AM modes. We have seen that it is possible to induce dual and anti-dual behaviours for certain incident beams $\Ei$, regardless of the nature of the particle. In this section, we quantify two hypothetical experiments that are feasible in the laboratory where the duality condition can be achieved. We will consider spheres made of Silica and Alumina. Their respective refractive indexes at $\lambda=780$nm are $\nsi=1.54$ \cite{Palik1985} and $\nal=1.76$ \cite{Bass1994}. We will suppose that they are embedded in water, therefore their respective relative refractive index will be $n_r^{\mathrm{SiO}_2}=1.16$ and $n_r^{\mathrm{Al}_2\mathrm{O}_3}=1.32$. 
 The way to proceed to induce helicity conservation will be the following. We consider the particle and its embedding medium as a given system. Then, we will see that we can achieve helicity conservation regardless of the nature of the particle (size and index of refraction), as long as it is approximately spherical, the surrounding medium is homogeneous, lossless and isotropic, and considering a tunable laser with a broad enough wavelength modulation. Once the two parameters $r$ and $n_r$ are known, we can always compute the range of size parameters $x$ that could be achieved with a tunable laser. Supposing that the tunable laser can offer wavelengths spanning from 700nm to 1000nm (that would be the case of a Ti:Sapphire laser, for example), the achievable $x$ will belong to the interval $\left\lbrace 6.28\, r(\mu \mathrm{m}), \ 8.98\, r(\mu \mathrm{m}) \right\rbrace$. Now, as it has been proven in the previous section, there also exist a large number of radii $r^*$ (and consequently, $x^*$) for which the dual condition is achieved. This is a consequence of the fact that given $n_r$, there exist different radius of particles $r_{m_z}$ that make the particle dual provided adequate excitation beams with angular momentum $m_z$ are used. Hence it is highly probable that regardless of $r$ and $n_r$ the dual condition $a_n=b_n$ can be achieved. In fact, this statement is even more true inasmuch as $r$ gets bigger. To be more specific, suppose that we have four different spheres. Two of them are made of Silica and the other two are made of Alumina. For each of the materials, suppose that the radius of the spheres are $r_1=325$nm and $r_2=700$nm. Now, given these sizes and their respective index of refraction, we can transform the $r$ dependence on the x axis in Fig. 3 and Fig. 5(b) into a $x=2\pi r / \lambda$ dependence and obtain the wavelength for which the dual condition will be achieved. The results are presented in Table I for the different four combinations $\left\lbrace r_1^{\mathrm{SiO}_2},r_2^{\mathrm{SiO}_2}, r_1^{\mathrm{Al}_2\mathrm{O}_3}, r_2^{\mathrm{Al}_2\mathrm{O}_3} \right\rbrace  $ of materials and sizes and for the two different excitation beams in consideration $m_z = 1$ and $5$ represented in Fig. 4. It can be observed that regardless of the size and the material, the duality condition can be achieved if a proper excitation beam is used. Moreover, we see that when the particle is larger, we need higher order beams to reach the duality condition, as long as we want to use visible wavelengths. Finally, it can also be observed that when the relative refractive index $n_r$ is increased, the duality condition is pushed to longer wavelengths. 
All this evidence makes us conclude that dual systems are easily realisable in the laboratory if an arbitrary dielectric sphere is properly matched with a proper light beam.

\begin{table}  
\begin{center}  
\begin{tabular}{| l | l | l |}
\hline  $\lambda$ (nm)& $m_z=1$ & $m_z=5$ \\
\hline $r_1^{\mathrm{SiO}_2}$ & \textbf{859} & 330 \\
\hline  $r_2^{\mathrm{SiO}_2}$ & 1860 & \textbf{710} \\
\hline $r_1^{\mathrm{Al}_2\mathrm{O}_3}$ & \textbf{986} & 377 \\
\hline  $r_2^{\mathrm{Al}_2\mathrm{O}_3}$ & 2122 & \textbf{812} \\
\hline 
\end{tabular}
\caption{Wavelengths at which the dual condition is achieved depending on the AM of the incident beam. The bold wavelengths are those at which the dual condition could be achieved with the the range of wavelengths available in a Ti:Sap laser. The dual conditions are achieved with a minimum precision of 2$\%$ for the four different cases.}
\label{table}
\end{center}
\end{table}

\section{Conclusions}
We have shown that the EM duality symmetry can be induced in dielectric spheres thanks to the use of higher AM modes. Our results show that the increase of relative index of refraction $n_r$ of the sphere and the order $m_z$ of the AM mode of the incoming beam help to achieve a dual sphere. Furthermore, we show the dependence of the radius of the particle with the dual condition. Finally, we show how flexible this method is by showing how to achieve helicity preservation with arbitrary dielectric particles.

\section{Acknowledgements}
This work was funded by the Australian Research Council Discovery Project DP110103697. G.M.-T. is the recipient of an Australian Research Council Future Fellowship (project number FT110100924).

\end{document}